\documentclass[prl,twocolumn,showpacs,amsmath]{revtex4}
\usepackage{epsfig}
\usepackage{bm}
\begin{document}
\title{Quantum computation with trapped polar molecules}
\author{D. DeMille}
\affiliation{Department of Physics, P.O. Box 208120, Yale
University, New Haven, CT 06520}
\date{October 27, 2001}
\begin{abstract}
We propose a novel physical realization of a quantum computer. The
qubits are electric dipole moments of ultracold diatomic
molecules, oriented along or against an external electric field.
Individual molecules are held in a 1-D trap array, with an
electric field gradient allowing spectroscopic addressing of each
site.  Bits are coupled via the electric dipole-dipole
interaction.  Using technologies similar to those already
demonstrated, this design can plausibly lead to a quantum computer
with $\gtrsim \! 10^4$ qubits, which can perform $\sim \! 10^5$
CNOT gates in the anticipated decoherence time of $\sim \! 5$ s.
\end{abstract}
 \maketitle
We describe a new technical approach to the design of a quantum
computer (QC).  The basic QC architecture is shown in Fig. 1.  The
qubits consist of the electric dipole moments of diatomic
molecules, oriented along or against an external electric field.
Bits are coupled by the electric dipole-dipole interaction.
Individual molecules are held in a 1-D trap array, with an
electric field gradient allowing spectroscopic addressing of each
site.  Loading with ultracold molecules makes it possible to use a
weak trapping potential, which should allow long decoherence times
for the system. This design bears various features in common with
other recent proposals which employ electric dipole couplings
\cite{Barenco95,Brennen99,Platzman99}. However, the technical
parameters of our design appear very favorable, and apparently
only incremental improvements of demonstrated techniques are
required in order to build a QC of unprecedented size.
\begin{figure}
\centerline{\psfig{figure=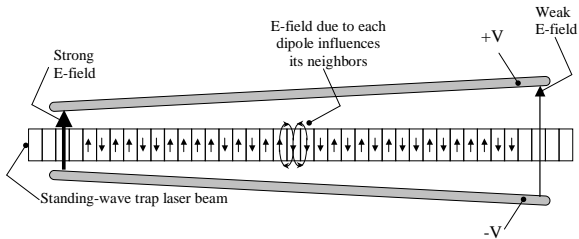,width=3.0 in}} \vspace{5 mm}
\caption{Schematic depiction of the polar molecule quantum
computer.  Molecules are trapped in a 1-D optical lattice.  Qubit
states correspond to electric dipole moments up or down relative
to the applied E-field. A field gradient makes the resonant
frequency for each qubit unique. The electric field of each dipole
changes the energy of its neighbors, according to their relative
orientations.} \label{Fig1}
\end{figure}

We describe the molecular qubits as permanent electric dipoles
oriented along ($\left| 0 \right\rangle $) or against ($\left| 1
\right\rangle $) an external electric field ($\vec E_{ext}$).
(This model reproduces the exact behavior well in a certain
regime.)  Lattice sites are equally spaced in the x-direction and
each contains one molecule, prepared initially in its ground state
$\left| 0 \right\rangle $. The external field is perpendicular to
the trap axis and consists of a constant bias field plus a linear
gradient: $\vec{E}_{ext}(x)=\left[ {E_0 + x \left( {\partial
E/\partial x} \right)} \right]\hat z$. The Hamiltonian for bit $a$
at position $x_a$ is $H'_a  = H^0  - \vec d_a \cdot \vec E_a$,
where $ H^0$ is the internal energy of a bit, $\vec d_a$ is the
electric dipole moment of bit $a$, and $\vec E_a = \vec E_{ext}
\left( {x_a } \right) + \vec E_{int} \left( {x_a } \right)$ is the
total electric field at $x_a$.  The internal field $\vec E_{int}$
is created by the electric dipole moments of neighboring bits:
$\vec E_{int} \left( {x_a } \right) = \sum\limits_{b \neq a}
{\frac{{- \vec d_b }}{{\left| {x_a - x_b } \right|^3 }}}$.  For
reasonable operating parameters, $E_{ext} \gg E_{int}$.

The scheme for gate operations is as outlined for the electric
dipole moments of quantum dots in Ref. \cite{Barenco95}.
Transitions between qubit states can be driven by electric
resonance, either directly in the microwave region or indirectly
by an optical stimulated Raman process.  Resonant drive pulses are
tuned to frequency $\nu _a  = \nu _0  + d_{eff}E_a /h$, where
$h\nu _0$ is the difference in internal energies between states
$\left| 0 \right\rangle$ and $\left| 1 \right\rangle$ in zero
field; the effective dipole moment $d_{eff} = |\vec d_{\left| 0
\right\rangle} - \vec d_{\left| 1 \right\rangle}|$, where $\vec
d_{\left| 0 \right\rangle (\left| 1 \right\rangle)}$ is the dipole
moment in state $\left| 0 \right\rangle (\left| 1 \right\rangle)$;
and $h$ is Planck's constant. Pulses of sufficient temporal length
to resolve the energy splitting due to $E_{int}$ can be used for
CNOT gates; shorter pulses suffice for one-bit rotations.
Final-state readout can be accomplished by state-selective,
resonant multiphoton ionization \cite{Demtroder96} and imaging
detection of the resulting ions and electrons.

The efficient creation of ultracold diatomic molecules by
photoassociation of laser cooled atoms was recently demonstrated
\cite{Fioretti98,Nikolov00,Drag00,Gabbanini00}. Electronically
excited neutral molecules are produced by a laser-induced
transition from the free state of two atoms; the excited state can
subsequently decay into bound vibrational levels of the molecular
ground state.  The molecules are formed at a translational
temperature similar to that of the constituent atoms; $T \approx
20~\mu \rm{K}$ has been demonstrated \cite{Drag00}.

Production of ultracold atoms is most advanced for alkali atoms.
Fortunately, heteronuclear bi-alkali molecules are well suited to
our purposes.  While no such species have yet been produced at
ultracold temperatures, there seems to be no fundamental obstacle
to making them.  The rate-limiting Franck-Condon (FC) factors in
the formation process in general should be more favorable for
hetero- than for homo-nuclear species, because of the better match
between ground- and excited-state potential curves
\cite{Wang98,Dion01}. Homonuclear bi-alkalis ${\rm K}_{\rm 2}$,
${\rm Rb}_{\rm 2}$, and ${\rm Cs}_{\rm 2}$ have been formed, as
well as heteronuclear molecular ions ${\rm NaCs}^{\rm +}$
\cite{Shaffer99}. Molecules formed by photoassociation are
typically in the lowest rotational states ($J=0-2$), but spread
over many vibrational levels ($v$). High vibrational levels ($v >
100$) of ${\rm Cs}_{\rm 2}$ were formed at a total rate of
$>10^6/{\rm s}$ \cite{Fioretti98,Drag00}; in a more complex
scheme, ${\rm K}_{\rm 2}$ molecules were produced at rates of up
to $10^5 /{\rm s}/{\rm level}$, in low vibrational states ($v \sim
10$) \cite{Nikolov00}. Based on the calculated FC factors and the
demonstrated production of homonuclear species, a production rate
of $\gtrsim \! 10^5/{\rm s}$ ultracold heteronuclear molecules in
individual rovibrational levels seems feasible. Molecules in any
state with $J=0$ or $2$ and $v \gg 1$ can be transferred
efficiently to the ground state ($v=0, J=0$) via a stimulated
Raman transition \cite{Gaubatz90}.

For the bi-alkali molecules, there is some tradeoff between ease
of production and the size of the molecular dipole moment. The FC
factors for photoassociation are largest for pairs of atoms with
similar excitation energies \cite{Wang98}, while the dipole
moments are largest for pairs where these are most different
\cite{Igel-Mann86}. We specifically consider the KCs molecule,
which has both a moderately large dipole moment and substantial FC
factors; however, the other bi-alkali species have similar
properties and one of them might prove ultimately more favorable.

An optical trap appears to be suitable for creating the desired
1-D array of molecules. For laser frequencies detuned to the red
of any electronic transition, the dynamic polarizability gives
rise to a force that attracts both atoms and molecules
\cite{Friedrich95} to regions of high intensity. Far off-resonance
traps are weak, but extremely non-perturbative \cite{Miller93}.
Such traps are well developed for atoms, with demonstrated trap
lifetimes $\gtrsim \! 300$ s \cite{OHara99}, and internal state
decoherence times $\gtrsim \! 4$ s \cite{Davidson95}. Trapping of
molecules in an off-resonant laser beam was recently demonstrated
for ultracold ${\rm Cs}_{\rm 2}$ \cite{Takekoshi98}.

Our proposed trap consists of a 1-D optical lattice, superposed
with a crossed dipole trap \cite{Adams95} of cylindrically focused
beams. This confines the molecules in sites spaced by $\lambda
_t/2$ (where $\lambda _t$ is the trap laser wavelength). The
molecules will be well localized in these wells for trap depth
$U_0 \gg kT$; we assume $U_0 = 100~\mu {\rm K}$ is sufficient.  We
take $\lambda _t \sim 1~\mu {\rm m}$ as a convenient compromise
between small trap spacing and increased decoherence rates.  For a
homogeneous trap of length $L$, we require that the Rayleigh
length $z_0  = \pi \omega _0^2 / \lambda _t > L$, where $\omega
_0$ is the beam waist.  We take $L = 5~{\rm mm}$ ($\sim \! 10^4$
trap sites) and $\omega _0 = 50~\mu {\rm m}$. Transverse
confinement is determined by the cylindrical beam waist $\omega _t
$; we assume diffraction-limited beams with f/1 focusing to
achieve $\omega _t \sim \lambda _t$.

For given $\lambda_t$ and laser power, the trap depth is
determined by the KCs dynamic polarizability, which is not known
in detail. However, it is possible to crudely estimate the
required parameters. For moderate laser frequency detuning
$\Delta$, the polarizability will be dominated by the oscillator
strength of the first excited $^1 \Sigma$ level, which should
couple to the ground state with a transition dipole moment
comparable to that for the $6s-6p$ transition of Cs
\cite{moleculardata}. For $\Delta \gg\ \! \omega_e$ (the molecular
vibrational frequency), the FC structure is irrelevant. Thus, for
the same detuning the trap depth for KCs should be similar to that
for atomic Cs. We find that $\Delta \approx 2000~cm^{-1}$ gives
reasonable behavior. For KCs this corresponds to a trap wavelength
$\lambda _t \approx 1.1~\mu {\rm m}$, and requires only $\approx
\! 1~{\rm W}$ of laser power for the 1-D lattice (as for Cs
\cite{Romalis99}). The cross-sectional area of each transverse
beam is $\sim2 \! \times$ that of the 1-D lattice beam, so the
power in these must be comparable to achieve transverse
confinement to $\sim \! \lambda _t /2$.  The required lasers are
commercially available.

K and Cs atoms can be loaded into such an optical trap from a
standard magneto-optic trap. If necessary, the temperature of the
atoms can be reduced in the trap by a variety of methods
polarization gradient cooling \cite{Winoto99, Adams95, Hamann98}.
The two-species sample in this trap should have $N \gtrsim 10^7$
atoms with density $n \gtrsim 10^{11}~{\rm cm}^{- 3}$ and $T
\lesssim 20~\mu {\rm K}$. Photoassociation for $\sim \! 1~{\rm s}$
and stimulated Raman transfer should produce $\sim \! 10^5$
molecules in the ground molecular state. Remaining atoms
(vibrationally-excited molecules) can be removed from the trap by
resonant light pushing (selective photoionization \cite{Dion01}).

Remarkably, it may prove relatively easy to distribute the
remaining molecules such that exactly one populates each lattice
site. It has been argued that the repulsive interaction between
atoms in a Bose condensate can lead to a Mott insulator-like phase
transition, and thus unity filling of an optical lattice
\cite{Jaksch98}. The interactions between polarized molecules are
many orders of magnitude stronger than for atoms, and thus may
facilitate reaching a similar phase transition even without Bose
condensation.  Detailed calculations are necessary to confirm this
speculation, which does not take into account the anisotropy of
the dipole-dipole interaction \cite{Santos00}.   The large
collision cross-sections for the polarized molecules \cite{Bohn01}
should also make it possible to achieve fast rates of evaporative
cooling, and thus (if necessary) an even lower temperature than
that of the original constituent atoms; the molecules can be held
in their ground state in this phase, to avoid losses due to
inelastic collisions.  We note in passing that the final molecular
temperature and density ($n \sim (2/\lambda_t)^3 \sim 10^{13}~{\rm
cm}^{- 3}$) discussed here correspond to a phase-space density of
$\sim \! 10^{-3}$, far from Bose condensation.

In the absence of an external field, even polar molecules have no
net electric dipole moment.  The application of an external field
mixes rotational states; for low fields the mixed state which
arises from the $J = 0~(J = 1, m_J  = 0)$ state corresponds to a
dipolar charge distribution along (against) $\vec E_{ext}$.
Calculations of the effect of $E_{ext}$ on these two states are
shown in Fig. 2.  The energies for $E_{ext} = 0$ are $E_J = hBJ(J
+ 1)$, where the rotational constant $B \approx 1.0~{\rm GHz}$ for
KCs \cite{Igel-Mann86}. Stark matrix elements are taken from
standard formulae \cite{Townes55}, using the calculated value of
the molecule-fixed dipole moment for KCs, $\mu = 1.92~ {\rm D}$
\cite{Igel-Mann86}.  In order to perform CNOT gates, it is
necessary to resolve the transitions $\left| 0 \right\rangle
\left| 0 \right\rangle \Leftrightarrow \left| 0 \right\rangle
\left| 1 \right\rangle $ from $\left| 1 \right\rangle \left| 0
\right\rangle \Leftrightarrow \left| 1 \right\rangle \left| 1
\right\rangle $. These differ in energy by $h\delta \nu  =
d_{eff}^2/\left( {\lambda_t/2} \right)^3$. Over a wide range of
electric field strengths $E_{ext} = (2 - 5)B/\mu$ ($\approx \!
2-5~{\rm kV/cm}$ for KCs), $d_{eff}$ is within $10\%$ of its
maximum value ($0.75~\mu $).  The time required for CNOT gates is
$\tau \gtrsim \left( {2\pi \delta \nu } \right)^{-1} \approx
50~\mu {\rm s}$. The one-bit drive frequencies $\nu _a $ cover the
range $3.5-6.0~{\rm GHz}$ over the array, with approximately equal
steps of $250~{\rm kHz}$ between sites. Direct microwave drive of
a CNOT gate requires rf electric field strength $\approx \!
10~{\rm mV/cm}$ for a $\pi $-pulse.
\begin{figure}
\centerline{\psfig{figure=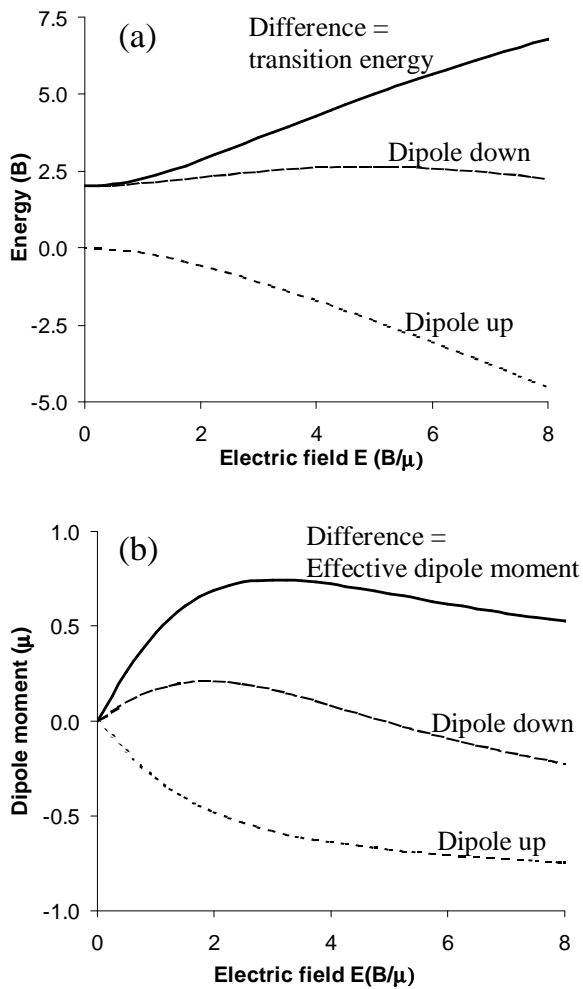,width=3.0 in}}\vspace{5 mm}
\caption{Effect of an electric field on a polar molecule.  a)
Energy levels.  b) Induced dipole moments.} \label{Fig2}
\end{figure}

The final state of the register can be read out by rapidly (but
adiabatically) turning off $\vec E_{ext}$, then applying a laser
pulse to perform resonantly-enhanced multiphoton ionization
\cite{moleculardata}.  Commercial pulsed lasers with $\sim \! {\rm
ns}$ pulse widths have both sufficient energy for $\sim \! 100 \%$
ionization efficiency ($\sim \! {\rm mJ/pulse}$), and sufficient
spectral resolution ($\ll \! 2B \approx 2~{\rm GHz}$) to make
contamination from the undesired logic state negligible. Molecules
in each state can be detected by consecutive identical laser
pulses, with an intervening rf $\pi$-pulse to transfer population
between logic states.  Simple ion optics can magnify the ionized
array image 10-fold, so that the charges form a pattern $5~{\rm
cm}$ long, with spacing between ions of $5~\mu {\rm m}$. The
magnified charge array can be detected on an imaging microchannel
plate. Commercial detectors are available with sufficient size and
resolution; detection of both ions and electrons from each logic
state should lead to effective efficiencies $\gtrsim \! 90\%$.

The most important known source of decoherence is photon
scattering from the trap laser. The total off-resonance photon
scattering rate is dominated by inelastic (Raman) scattering to
other rotational and vibrational levels \cite{Loudon83}.  For the
chosen value of $\Delta$, the scattering rate for KCs should be
comparable to the elastic scattering rate $R_s$ for Cs (much as
for the trap depth). For the trap parameters discussed, $R_s \sim
0.2~{\rm s}^{-1}$ \cite{Romalis99}.

We have considered several technical noise issues, all of which
appear controllable at the desired level. The trap laser shifts
the values of $\nu_a$, through coupling to the tensor
polarizability of the molecule. Tensor shifts are typically
several times smaller than the scalar shifts ($U_0 \approx 2~{\rm
MHz}$) responsible for the trapping potential \cite{Friedrich95};
we conservatively assume a tensor shift as large as $U_0$. We
require that the 1-bit drive frequency have noise $\delta \nu _a
\lesssim \sqrt{R_s} \sim 0.5~{\rm Hz}/\sqrt{\rm Hz}$
\cite{Lamoreaux97}; this implies laser intensity stability $\delta
I / I \lesssim 3\times 10^{-7} / \sqrt{\rm Hz}$. This is $\sim \!
300 \times$ the shot-noise limit, and should be achievable
\cite{Hall86}. Electric field noise couples directly to the
molecular dipole moments, and is also of concern. With field plate
spacing of $\sim \! 1~{\rm cm}$, we require broadband voltage
noise $\delta V \lesssim 0.5~\mu {\rm V}/\sqrt{\rm Hz}$, the
room-temperature Johnson noise on a $10~{\rm M} \Omega$ resistor.
Noise from the high-voltage supply can be heavily filtered and
should pose no problems. A variety of other possible decoherence
sources seem to present no limitations. These include heating due
to laser intensity, beam-pointing, or frequency fluctuations
\cite{OHara99,Savard97}; dissociation of molecules by the trap
laser \cite{Askaryan65}; spontaneous emission; coupling to
blackbody radiation; collisions with background gas molecules;
etc.

We have shown that a quantum computer based on ultracold KCs
molecules can plausibly achieve $\sim \! 10^5$ CNOT gates on $\sim
\! 10^4 $ bits in the anticipated decoherence time of $\sim \!
5~{\rm s}$.  This may be sufficient for quantum error correction
methods to ensure that arbitrarily long computations are stable
\cite{Preskill98}. We have also argued that this system requires
no dramatic technical breakthroughs for its initial construction.
The electric resonance techniques for the processor should be
robust and easy to implement, by analogy with similar NMR methods.
Creation of the trapped array of polar molecules appears to be a
direct extension of recent work in laser cooling and trapping, and
the readout via resonance-enhanced ionization is standard.  Unlike
recent proposals for quantum computation using ultracold atoms,
our technique requires neither mechanical motion
\cite{Brennen99,Jaksch99},  nor coupling to short-lived excited
states \cite{Brennen99,Jaksch00}, for gate operations.

There are a number of potentially serious issues that we have not
considered.  For example, the bit-bit interaction cannot be
switched off, and thus operation will require techniques similar
to the "refocusing" procedure used to control the couplings in NMR
quantum computation \cite{Nielsen00}. We have ignored the motional
states of the molecules; although the trap motional frequencies
($\sim \! 100~{\rm kHz}$) are well-separated from other frequency
scales in the device, couplings of gate operations to the motion
may cause additional decoherence or gate fidelity loss
\cite{Jaksch00}. We have also ignored the hyperfine structure of
the KCs molecules, which might complicate the initial state
selection and/or gate operations.  We plan to investigate these
issues in the future.  In the meantime, we have begun an
experimental effort to implement these ideas (using RbCs rather
than KCs for technical convenience).

On the other hand, the parameters discussed here might also be
improved with other techniques that are currently less well
developed.  For example, buffer-gas cooling \cite{Weinstein98} or
electric slowing and trapping \cite{Bethlem00}, in combination
with evaporative cooling \cite{Doyle95}, could yield larger and/or
colder samples; the variety of molecules accessible to these
techniques could enable the use of larger values of $\mu$ and/or
smaller values of $\lambda_t$. Microfabricated traps based on
low-frequency electromagnetic fields might prove advantageous
\cite{Calarco00}, and non-destructive readout may be possible by
direct pickup of the molecular dipole fields with nearby
single-electron transistors \cite{Schoelkopf98}. Finally, in
addition to our qubit states, there are $\sim \! 10^6$ long-lived
rovibrational states available for each molecule
\cite{moleculardata}; these might allow each molecule to function
as a quantum information unit containing $n \gg 1$ bits of
information. Although entanglement between individual molecules is
more difficult in this case, the massive parallelism involved may
be useful in itself \cite{Ahn00}.

We thank M. Kasevich, P. Zoller, and A.J. Kerman for useful
discussions.  DD is an Alfred P. Sloan Research Fellow and a
Packard Foundation Fellow.  This work is supported by NSF ITR
grant \#EIA-0081332.

\end{document}